\documentclass[lettersize,journal]{IEEEtran}
\usepackage{amsmath,amsfonts}
\usepackage{algorithmic}
\usepackage{algorithm}
\usepackage{amsmath,amssymb,amsfonts}
\usepackage{array}
\usepackage[caption=false,font=normalsize,labelfont=sf,textfont=sf]{subfig}
\usepackage{textcomp}
\usepackage{stfloats}
\usepackage{url}
\usepackage{verbatim}
\usepackage{graphicx}
\usepackage{booktabs}
\usepackage{cite}
\usepackage{makecell}
\usepackage{multirow}
\usepackage{bm}
\usepackage[justification=centering]{caption}
\usepackage{adjustbox}
\usepackage[dvipsnames,usenames]{color}
\usepackage{mathrsfs}
\usepackage[table]{xcolor}
\begin{document}
	
	\title{ConsisFormer: Compute-Efficient Transformer \\for Wireless Foundation Models Based on \\Channel Consistency}
	
	\author{
		\IEEEauthorblockN{Yuwei~Wang,~\IEEEmembership{Member,~IEEE,}
			Li~Sun$^\ast$,~\IEEEmembership{Senior Member,~IEEE,}
			Tingting~Yang,~\IEEEmembership{Senior Member,~IEEE,}
			Liwen~Jing,~
			Yuxuan~Shi,~
			Maged~Elkashlan,~\IEEEmembership{Senior Member,~IEEE,}
			Mérouane~Debbah,~\IEEEmembership{Fellow,~IEEE}
			}
		\thanks{
		Yuwei Wang, Li Sun, Tingting Yang, Liwen Jing, and Yuxuan Shi are with the Department of Broadband Communication, Pengcheng Laboratory, Shenzhen, 518000, China (e-mail: wangyw03@pcl.ac.cn; sunl03@pcl.ac.cn; yangtt@pcl.ac.cn, jinglw@pcl.ac.cn, shiyx01@pcl.ac.cn, Corresponding Author: Li Sun).
		
		Maged Elkashlan is with the School of Electronic Engineering and Computer Science, Queen Mary University of London, E1 4NS London, U.K. (e-mail: maged.elkashlan@qmul.ac.uk).
		
		Mérouane Debbah is with the Department of Computer and Information Engineering, Khalifa University of Science and Technology, Abu Dhabi, 127788, UAE (e-mail: merouane.debbah@ku.ac.ae).	
}} % <-this % stops a space% <-this % stops a space
	% <-this % stops a space
	
	
	%\author{
		%\normalsize Yuwei Wang$^\dagger$, Li Sun$^\dagger$
		%
		%\\  $^\dagger$ Department of Network Intelligence, Pengcheng Laboratory, Shenzhen, China
		%\\ Emails: \{\textit{wangyw03@pcl.ac.cn}, \textit{sunl03@pcl.ac.cn}\}}

	\maketitle
	\thispagestyle{empty}
	\begin{abstract}
Wireless foundation models (WFMs) have recently emerged as a promising paradigm for AI-native 6G networks, enabling universal channel representations adaptable to diverse communication and sensing tasks. Existing WFMs are predominantly built upon the Transformer architecture, which delivers superior performance but incurs computational complexity proportional to the square of the input sequence length, posing a significant barrier to their deployment under stringent inference latency constraints. To address this issue, in this paper, we propose ConsisFormer, a compute-efficient Transformer design based on short-term consistency of wireless channels, as a WFM backbone. By utilizing the observation that adjacent time or frequency instances share similar clusters of scatterers and thus exhibit similar channel characteristics, we develop an adaptive token aggregation (ATA) module to dynamically merge neighboring channel state information (CSI) tokens, thereby reducing the length of the token sequence involved in self-attention calculations to lower the computational cost. Furthermore, we propose a feature sequence interpolation (FSI) method to recover the full CSI representation based on the sparse feature sequence outputted from the Transformer blocks, thus keeping the performance unaffected while ensuring low complexity. Moreover, we propose an aggregated auto-encoder (AAE) pre-training paradigm for WFMs, enabling robust channel representation learning from sparsified CSI tokens via compression and recovery. Simulation results show that the proposed design reduces the computational complexity of WFM by over $83\%$ with negligible performance loss on various tasks including channel prediction, LoS/NLOS classification, beam prediction, and localization.
	\end{abstract}
	
	
	\vspace{-16pt}
\section{Introduction}
In its IMT-2030 (6G) vision published in 2023, ITU-R clearly identified the integration of AI and communications as one of the six pillars for 6G \cite{ITU}. AI models deployed at the physical layer (PHY) of radio access networks (RAN) can be utilized to better support channel modeling and prediction, enable efficient time-frequency resource scheduling, improve spectral efficiency, and empower advanced transceiver design, among other things \cite{Jun}-\cite{JSAC}.

The integration of AI and communications has gone through the 1.0 era and is now entering the 2.0 era. The ``AI $+$ communications" 1.0 era was primarily based on using task-customized small AI models to perform modular optimization or end-to-end reconstruction of communication links. In this era, task-specific Transformer-based models have been widely adopted for various physical layer tasks such as channel prediction, beamforming, signal detection, and power optimization \cite{Wang}-\!\!\cite{Chen}, but generally suffer from limited generalization capability and high retraining overhead across diverse tasks and scenarios. Since the birth of ChatGPT at the end of 2022, Large Language Models (LLMs) have already demonstrated astonishing capabilities in areas such as content generation, sequence modeling, and conversational interaction. Correspondingly, researchers in the communications field have shown increasing interests in building wireless foundation models (WFMs), thus propelling ``AI $+$ communications" into the 2.0 era. In the ``AI $+$ communications" 2.0 paradigm, a task-agnostic foundation model is first pre-trained on large-scale datasets to learn universal feature representations of the wireless channels. Then, techniques such as fine-tuning are used to adapt the unified representations to downstream task models or algorithms, thereby achieving generalization across tasks and scenarios. There are two major technical paths for building the WFMs. 

The first path is based on fine-tuning the open-source LLMs or multi-modal large models. In particular, \cite{Liu} proposed a channel predictor LLM4CP, where the Add-and-Norm and output layers of GPT-2 were fine-tuned to predict downlink channels based on historical uplink channels. In \cite{Yucheng}, the wireless time series data was reprogrammed into a natural language representation and aligned with the pre-training input format of LLMs, such that the ability of LLMs in sequence modeling can be exploited to realize beam prediction. Instead of using LLMs, \cite{Yizhu} addressed the beam prediction problem by fine-tuning the multi-modal large model DeepSeek, where positions of the scatters and multi-view images were utilized to extract environmental embeddings and facilitate the generation of optimal beam index. 

The second path is developing wireless native foundation models by training from scratch. In \cite{Boxun}, a masked auto-encoder (MAE) based network structure was devised to build a WFM for channel prediction in both time domain and frequency domain, where extensive CSI datasets with diversified configurations were utilized for pre-training. In \cite{Feifei2}, a decoder-only architecture was proposed to further enhance the performance of channel prediction, where both next channel prediction (NCP) loss and masked channel reconstruction (MCR) loss were used such that the model better captures the underlying characteristics of wireless channels. While the advantages of building WFMs over small models have been demonstrated in \cite{Liu}-\!\!\cite{Feifei2}, these works only focus on a single type of task. In contrast, \cite{Sadjad} developed a task-agnostic model called Large Wireless Model (LWM) to generate universal channel embeddings that can be used across a wide range of downstream communication tasks. Following this paradigm, subsequent works have further enhanced the generality of WFMs from different perspectives. For example, \cite{Yang} incorporated cross-domain embeddings into the Transformer architecture, enabling a unified model to support both communication and sensing tasks. Similarly, \cite{Liwen} proposed a self-supervised pre-trained WFM that can be applied to CSI feedback and human activity recognition. In \cite{Yucheng1}, a granularity encoding is introduced to distinguish different types of tasks, enabling the model to adaptively extract the most relevant channel features corresponding to various tasks. To enhance cross-scenario generalization, CSI-CLIP \cite{Shugong} employs contrastive learning on paired CSI and CIR data to learn more robust wireless representations. Furthermore, MUSE-FM \cite{MUSE} introduces a prompt-guided encoder-decoder and environmental contexts to handle heterogeneous tasks and cross-scenarios adaptation. 

Despite the differences in model details and pre-training approaches, all of the WFMs proposed so far such as \cite{Liu}-\cite{Yucheng1} use the Transformer architecture as their backbones, which is able to capture both local and global features of the CSI sequence to produce unified channel representations. In the context of physical layer communications, the Transformer-based neural networks typically accept a sequence of CSI samples (tokens) as inputs, which are processed by multiple Transformer blocks stacked together with each containing self-attention layers and fully-connected layers. Note that the computational complexity of self-attention is proportional to the square of the input sequence length \cite{Vaswani}. This property is not favorable for many wireless applications where long sequences consisting of multiple CSI samples as tokens have to be processed. Thus, the application of Transformers in practical systems is challenging due to the stringent constraints on inference latency, especially for time-sensitive scenarios (e.g., channel estimation, signal detection, and channel prediction).

To deal with this issue, in this paper, we propose \textbf{ConsisFormer}, a channel-\underline{\textbf{consis}}tency driven Trans\underline{\textbf{former}} structure, to realize compute-efficient self-attention calculation and channel feature extraction. Channel consistency\footnote{The channel consistency can be interpreted from the time, frequency, and space domains. The frequency domain consistency indicates that closely spaced subcarriers are often highly correlated, while the space domain consistency refers to the phenomenon that adjacent antenna elements sharing similar positions have many identical scatterers/clusters. For ease of presentation, the proposed method is described based on temporal channel consistency, while the same principle is applicable to the frequency and space domains as well.} is an important property of wireless channels whereby adjacent time instances share similar propagation environments and thus exhibit similar channel characteristics. Based on this property, not all CSI tokens\footnote{Each CSI token corresponds to a short segment of CSI samples.} are necessary for attention calculation, and it suffices to retain only representative tokens to capture the intrinsic features of the wireless channels. Keeping this in mind, we propose an adaptive token aggregation (ATA) method to progressively merge neighboring CSI tokens by utilizing their similarity, significantly reducing the number of tokens involved in self-attention calculations. Although token sparsification offers a compute-efficient solution, it may incur performance degradation due to the loss of information in the discarded tokens. To overcome this, we further develop a feature sequence interpolation (FSI) module, which is deployed after the Transformer blocks. This module produces a full-length feature sequence based on the sparse feature vectors outputted by the Transformer blocks, thereby compensating for the loss in representation capability. Together, the ATA and FSI modules form a principled aggregation–extraction–recovery design paradigm, where channel information is first compressed, then extracted by the Transformer blocks, and finally recovered for downstream task inference.

Building upon this paradigm, we further propose a novel pre-training strategy, termed as aggregated auto-encoder (AAE), to train ConsisFormer as a WFM backbone. Existing MAE-based pre-training randomly masks CSI tokens, which may discard critical channel components and leave the remaining observations insufficient to characterize the underlying propagation structure. In contrast, AAE replaces random masking with channel-consistency-based structured aggregation, where the removed information mainly corresponds to redundant or highly correlated CSI components. As a result, the model is encouraged to recover complete CSI representations from compressed yet physically meaningful observations, leading to more robust and effective pre-training. Combined with AAE, ConsisFormer achieves significant computational complexity reduction while preserving strong task performance, making it a high-performance and compute-efficient backbone for WFMs.

While existing efforts toward efficient Transformers in wireless communications have explored model compression techniques such as pruning, quantization, and knowledge distillation \cite{DHou}-\!\!\cite{Byeon}, and token reduction methods have also been studied in other fields such as computer vision \cite{Bolya}-\!\!\cite{JYun}, these approaches are either parameter-level optimizations or domain-agnostic designs that do not leverage the physical properties of wireless channels. In contrast, ConsisFormer is, to the best of our knowledge, the first to exploit channel consistency for token-level sparsification in Transformer-based models, with the aggregation–extraction–recovery paradigm offering a novel and domain-specific path toward compute-efficient Transformer design. The major contributions of this paper are summarized as follows.

\begin{itemize}
	\item First, we propose a channel-consistency inspired adaptive aggregation method to dynamically merge neighboring CSI tokens with intrinsic similarities. Via multi-round iterations, the aggregation algorithm is able to produce a shorter CSI token sequence, which is processed by a series of Transformer blocks with each containing self-attention and FFN layers, thus significantly reducing the computational complexity of the Transformer.
	\item Second, we develop a correlation-based interpolation module to compensate for the possible performance degradation incurred by the compute-efficient design. In particular, CSI features produced by the final Transformer block are first encoded into a correlation representation. Given the correlation representation, the original CSI tokens, and the indices of tokens, a query vector is generated, which is used to interpolate the available CSI features to form a fine-grained representation of the wireless channels.
	\item Third, based on the proposed ConsisFormer structure, we introduce a novel pre-training paradigm, termed as aggregated auto-encoder (AAE), for building WFMs. The AAE incorporates information compression and recovery into the pre-training objective, encouraging the model to learn informative and robust channel representations from sparsified token sequences.

\end{itemize}

The rest of the paper is organized as follows. The system model and existing structure are shown in Section II. Then the ConsisFormer design is presented in Section III, where the adaptive aggregation algorithm and the correlation-based interpolation approach are described in turn. Afterwards, the aggregated auto-Encoder pre-training method is presented in Section IV. Simulation results with performance comparisons are given in Section V. Finally, we conclude this paper in Section VI.
	
\section{System Model and Existing Structure} 

\subsection{System Model}
In this paper, we consider a MIMO-OFDM communications system with $N$ transmit antennas and $M$ receive antennas. Each slot consists of $T$ OFDM symbols in the time domain and $K$ subcarriers in the frequency domain. Within a slot, the whole channel matrix over all REs is $\bm{H}\in\mathbb{C}^{T\times K\times S}$, where $S=M\cdot N$ is the number of spatial channels. The wireless channel can be characterized by a multipath propagation model. Specifically, the frequency-domain MIMO channel at the $t$-th OFDM symbol and $k$-th sub-carrier can be expressed as
\begin{eqnarray}
	\bm{H}[t,k]=\sum_{\ell=1}^{N_L}\alpha_\ell[t]\bm{a}_r(\theta_\ell)\bm{a}_t^{\mathrm{H}}(\phi_\ell)e^{-j2\pi f_k\tau_\ell[t]},
\end{eqnarray}
where $N_L$ is the number of paths, $\alpha_\ell[t]$ and $\tau_\ell[t]$ denote the complex gain and delay of the $\ell$-th path, respectively, $\theta_\ell$ and $\phi_\ell$ represent the angle-of-arrival (AoA) and angle-of-departure (AoD), and $\bm{a}_r(\cdot)$ and $\bm{a}_t(\cdot)$ are the array response vectors \cite{PDong}.

\begin{figure}[t]
	\begin{center}
		\includegraphics[scale=0.5]{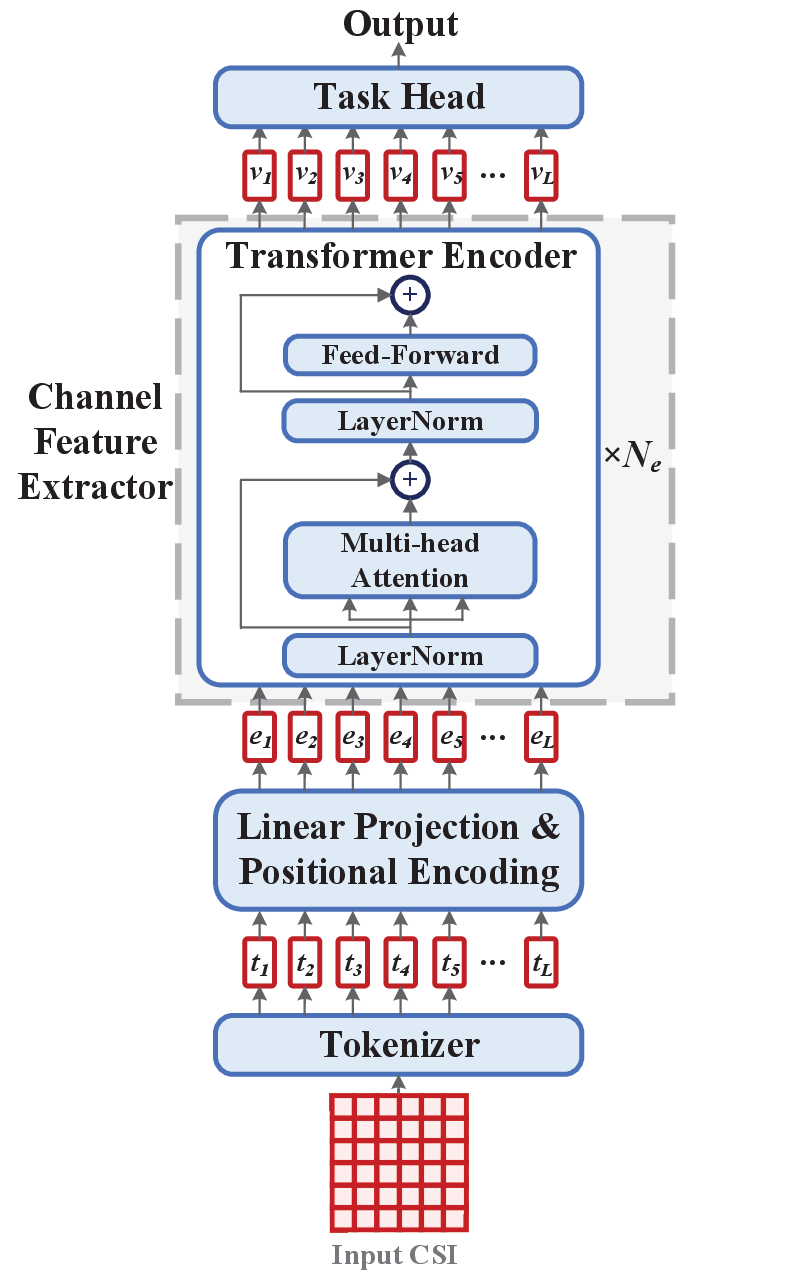}
		\captionsetup{justification=justified, singlelinecheck=false}
		\caption{\small{The existing classic Transformer-based model for WFMs. The input CSI matrix is segmented into $L$ tokens by the tokenizer, projected into embeddings with positional encoding, and then processed by $N_e$ Transformer encoder blocks. The output features are passed to task heads for downstream tasks.}}
	\end{center}
	\label{fig1}
	\vspace{-19pt}
\end{figure}

This model reveals that the channel response is jointly determined by a set of slowly-varying physical path parameters, as $\alpha_\ell[t]$ and $\tau_\ell[t]$ vary slowly over time while $e^{-j2\pi f_k\tau_\ell[t]}$ changes smoothly across sub-carriers \cite{PBello}. As a result, adjacent resource elements tend to share similar channel characteristics, and the corresponding CSI tokens contain redundant information.

\subsection{Existing Transformer-Based Structure}

The existing structure of Transformer-based WFMs is illustrated in Fig. 1. As is exhibited in Fig. 1, the channel matrix first goes through a tokenizer, which reshapes and segments the input into a sequence of tokens $\bm{t}_1, \bm{t}_2, ..., \bm{t}_L$, where the length of the token sequence is $L$. Each token $\bm{t}_l (1\leqslant l \leqslant L)$ is then fed into a linear projection and positional encoding module to produce a sequence of $d$-dimensional embedding vectors as
\begin{eqnarray}
	\bm{e}_l=\bm{W}_{\text{proj}}\bm{t}_l+\mathcal{P}(l),
\end{eqnarray}
where $\bm{W}_{\text{proj}}$ is linear projection matrix, and $\mathcal{P}(\cdot)$ denotes the positional encoding function. Each embedding $\bm{e}_l (1\leqslant l \leqslant L)$ is a vector of dimension $d$ that jointly captures the channel characteristics and their positional information. Subsequently, the embedding sequence is passed through a series of Transformer encoder blocks, which acts as a channel feature extractor and outputs a sequence of feature vectors $\bm{v}_l (1\leqslant l \leqslant L)$. These feature vectors are further processed by a task head to generate the desired outputs. For instance, an extrapolation head, a regression head, and a classification head are employed for channel prediction, localization, and beam prediction tasks, respectively. Note that the architecture in Fig. 1 represents a typical existing design, which serves as the baseline for subsequent analysis.

\subsection{Computational Complexity Analysis}
In the aforementioned architecture, the dominant computational cost arises from the Transformer encoder blocks. Specifically, for each block, the input embedding vectors are first transformed to query, key, and value vectors, and the matrix representation of this process is 
\begin{eqnarray}
	\bm{Q}^{\text{att}}=\bm{W}_q\bm{X},~\bm{K}^{\text{att}}=\bm{W}_k\bm{X},~\bm{V}^{\text{att}}=\bm{W}_v\bm{X},
\end{eqnarray}
where $\bm{X}\in\mathbb{R}^{d\times L}$ denotes the input of the current Transformer block, and $\bm{W}_q\in\mathbb{R}^{d\times d}$, $\bm{W}_k\in\mathbb{R}^{d\times d}$, and $\bm{W}_v\in\mathbb{R}^{d\times d}$ are projection matrices of the current Transformer block. Since the dimensions of query, key, and value vectors are equal to that of the input embedding vector, the computational complexity of linear transform is $\mathcal{O}(Ld^2)$. Afterwards, the multi-head attention layer computes the correlations between query vectors and key vectors, based on which the weighted sum of value vectors is obtained as 
\begin{eqnarray}
	\bm{O}^{\text{att}}=\bm{V}^{\text{att}}\sigma(\dfrac{(\bm{Q}^{\text{att}})^T\bm{K}^{\text{att}}}{\sqrt{d}})\in\mathbb{R}^{d\times L}.
\end{eqnarray}
In (4), $\sigma(\cdot)$ is the column-wise softmax function, and $(\cdot)^T$ is the transpose operation. According to (4), the computational complexity of multi-head attention layer is $\mathcal{O}(L^2d)$, which scales quadratically with the token sequence length $L$. 

\begin{figure*}[t]
	\begin{center}
		\includegraphics[scale=0.54]{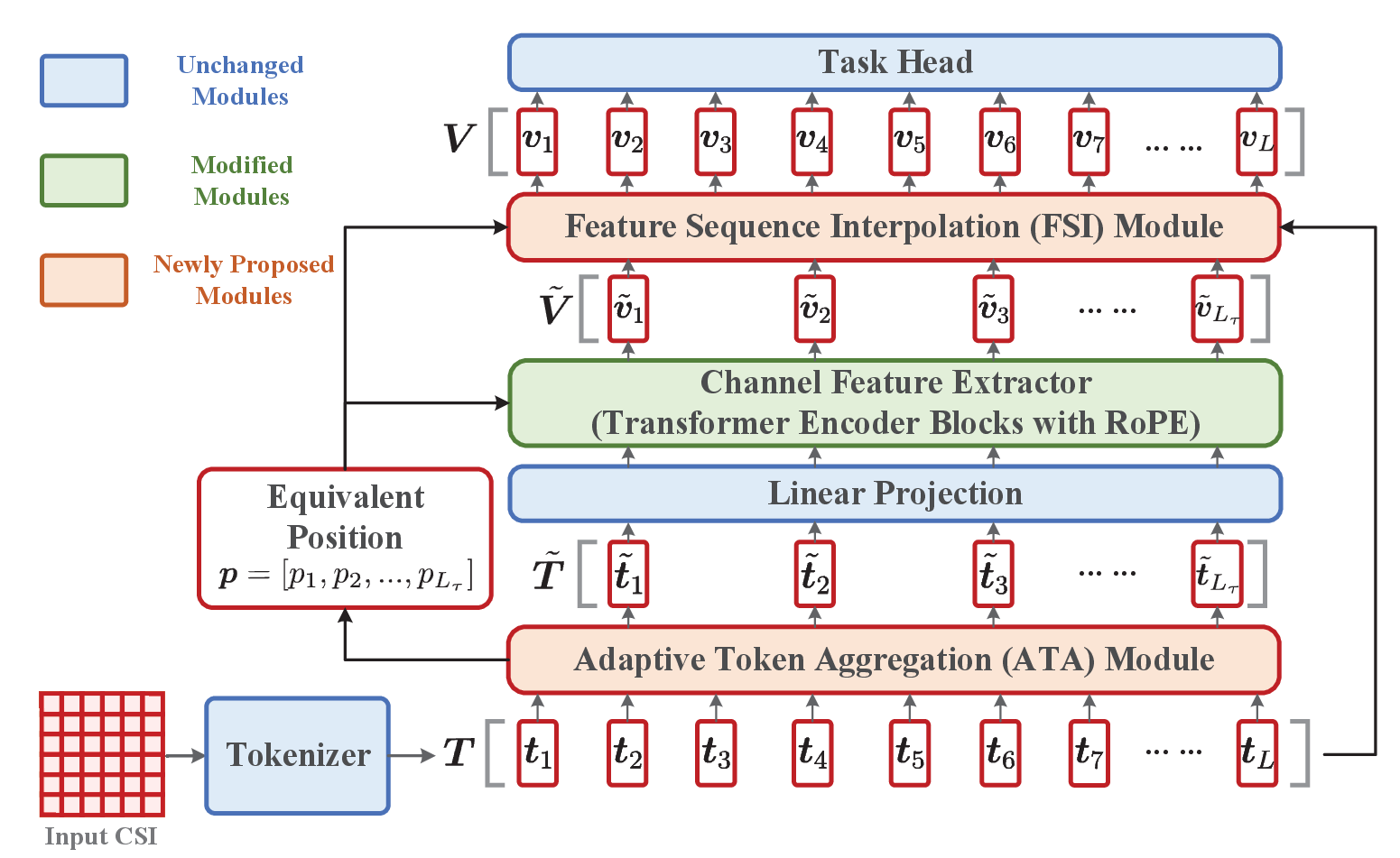}
		\captionsetup{justification=justified, singlelinecheck=false}
		\caption{\small{The structure of the proposed ConsisFormer. The original token sequence $\bm{T}$ is first aggregated into a shorter sequence $\tilde{\bm{T}}$ of length $L_\tau$ by the ATA module, along with equivalent positions $\bm{p}$. The aggregated tokens are then fed into the Transformer encoder blocks with RoPE to extract sparse channel features $\tilde{\bm{V}}$. Finally, the FSI module recovers the full-length feature sequence $\bm{V}$ for task heads.}}
	\end{center}
	\label{fig2}
	\vspace{-16pt}
\end{figure*}

After a residual connection and a layer normalization, each vector $\bm{o}_l\in\mathbb{R}^{d\times 1} (1\leqslant l \leqslant L)$ in $\bm{O}^{\text{att}}$ is fed into a feed-forward network (FFN). FNN consists of a linear dimension expansion layer with expand coefficient $\alpha$ and a linear dimension reduction layer that scales the output dimension back to $d$ as
\begin{eqnarray}
	\text{FFN}(\bm{o}_l)=\bm{W}_2\sigma(\bm{W}_1\bm{o}_l),
\end{eqnarray}
where $\bm{W}_1\in\mathbb{R}^{\alpha d\times d}$ and $\bm{W}_2\in\mathbb{R}^{ d\times  \alpha d}$ are the parameter matrices. Therefore, the computational complexity of FFN is $\mathcal{O}(\alpha Ld^2)$.

Assuming the number of Transformer encoder blocks as $N_e$, the overall computational complexity of the channel feature extractor is $\mathcal{O}(N_e\alpha Ld^2+N_e L^2d)$, which is highly sensitive to the token sequence length $L$. In next-generation communications systems, the use of wider bandwidths and larger antenna arrays significantly increases the dimension of the channel matrix $\bm{H}$ and thus enlarge the token sequence length $L$. Consequently, the computational cost of Transformer-based channel representation grows rapidly, posing a major challenge for real-time physical layer deployment, especially for low-latency scenarios.

\section{Compute-Efficient Transformer Design}

As discussed in Section II, the computational complexity of Transformer-based channel representations increases rapidly with the length of the token sequence $L$. However, in practical wireless propagation environments, CSI exhibits strong consistency and redundancy across time and frequency domains, as adjacent OFDM symbols or subcarriers often carry highly correlated channel characteristics representing the similar underlying physical environment. This inherent redundancy implies that feeding all tokens into the Transformer encoder is unnecessary and computationally inefficient. Motivated by this observation, we propose a compute-efficient Transformer architecture termed as ConsisFormer, as illustrated in Fig. 2.

\subsection{Overall Structure of ConsisFormer}

Compared with the structure in Fig. 1, the proposed ConsisFormer architecture consists of three unchanged modules, one modification of existing modules, and two newly proposed modules, which are represented by blue, green, and orange blocks, respectively. 

In our proposed architecture, the channel matrix is first processed by a tokenizer, which reshapes and partitions the input channel into a sequence of tokens along the time, frequency, or space dimension. For clarity of presentation, the proposed methods are described assuming a time-domain tokenization in this section. Nevertheless, the same design principles and processing procedures are directly applicable to frequency-domain and space-domain tokenization. In the tokenizer, the channel matrix $\bm{H}\in\mathbb{C}^{T\times K\times S}$ is flattened along frequency and spatial dimensions, and then segmented to a sequence of tokens $\bm{T} = [\bm{t}_1, \bm{t}_2, ..., \bm{t}_L]$, where $\bm{t}_l\in\mathbb{C}^{KS\times1} (1\leqslant l \leqslant L)$. 

Afterwards, the full token sequence is fed into an adaptive token aggregation (ATA) module to generate an aggregated token sequence by exploiting channel consistency, which eliminates redundant information while preserving the essential correlation and structural characteristics of the CSI. The design of this module will be detailed in Section III-B. With $\bm{T}$ as the input, the ATA module produces an aggregated token sequence with length $L_\tau (L_\tau \leqslant L)$ and the corresponding equivalent positions as
\begin{eqnarray}
	(\tilde{\bm{T}}, \bm{p})=\mathcal{A}(\bm{T}),
\end{eqnarray}
where $\mathcal{A}(\cdot)$ represents the aggregation module, $\tilde{\bm{T}} = [\tilde{\bm{t}}_1, \tilde{\bm{t}}_2, ..., \tilde{\bm{t}}_{L_\tau}]$ is the aggregated token sequence, and $\bm{p}=[p_1, p_2, ..., p_{L_\tau}]$ is the vector consisting of equivalent positions. In the ATA module, the tokens are merged dynamically, which results in a non-uniform aggregation across the original sequence. Consequently, the aggregated tokens do not have uniformly sampled positions any longer, and their indices cannot accurately represent the relative positional relationships among tokens. To address this, equivalent positional information is jointly generated during token aggregation, enabling the Transformer encoder to correctly capture the structural dependencies among aggregated tokens.

The aggregated complex tokens $\tilde{\bm{t}}_l\in\mathbb{C}^{KS\times 1} (1\leqslant l \leqslant L_\tau)$ are reshaped and projected to real-valued embeddings with the dimensions of $d$, i.e., $\tilde{\bm{e}}_l\in\mathbb{R}^{d\times 1} (1\leqslant l \leqslant L_\tau)$, which are fed into the channel feature extractor along with the equivalent positions to obtain a feature sequence
\begin{eqnarray}
	\tilde{\bm{V}}=\mathcal{E}(\tilde{\bm{T}}, \bm{p}),
\end{eqnarray}
with $\tilde{\bm{V}}=[\tilde{\bm{v}}_1, \tilde{\bm{v}}_2, ..., \tilde{\bm{v}}_{L_\tau}]$ being a sequence of $L_\tau$ feature vectors. The channel feature extractor is composed of Transformer encoder blocks with rotary position embedding (RoPE), whose details are explained in Section III-C. Owing to the short-term consistency of wireless channel, in most practical scenarios (except for extremely high-mobility conditions), the length of aggregated token sequence $L_\tau$ is significantly smaller than that of the original token sequence $L$. Consequently, the proposed token sparsification method substantially reduces the overall computational cost.

Afterwards, the feature sequence $\tilde{\bm{V}}$, which is produced based on aggregated token sequence $\tilde{\bm{T}}$, goes through a feature sequence interpolation (FSI) module along with the equivalent positions and the original tokens $\bm{t}_l (1\leqslant l \leqslant L)$ to generate a full-length feature sequence
\begin{eqnarray}
	\bm{V}=\mathcal{I}(\tilde{\bm{V}}, \bm{p}, \bm{T}),
\end{eqnarray}
where $\bm{V}=[\bm{v}_1, \bm{v}_2, ..., \bm{v}_{L}]$. As will be illustrated in Section III-D, although the original tokens are utilized in the FSI module, they are processed independently without mutual interactions. In addition, the FSI module is applied only at the output of the final layer of the channel feature extractor. As a result, the computational overhead introduced by the FSI module is negligible. 

The motivation for introducing the FSI module is twofold. First, during the ATA process, a certain degree of information loss is inevitable, and the proposed FSI module can be used to compensate for the resulting performance degradation by leveraging the extracted features and the original token information. Second, the length of sparse token sequence $L_\tau$ varies for different input CSI data, whereas the subsequent task head may not be able to accommodate variable-length inputs. Therefore, the feature sequence needs to be completed to the full length, ensuring compatibility and generalization capability with the task head. The design of FSI module is elaborated on in Section III-D. 

After the FSI module, the complete feature vectors $\bm{v}_l (1\leqslant l\leqslant L)$ are processed by a task head to produce the final result for the specific communications or sensing tasks. 

\begin{figure*}[t]
	\begin{center}
		\includegraphics[scale=0.65]{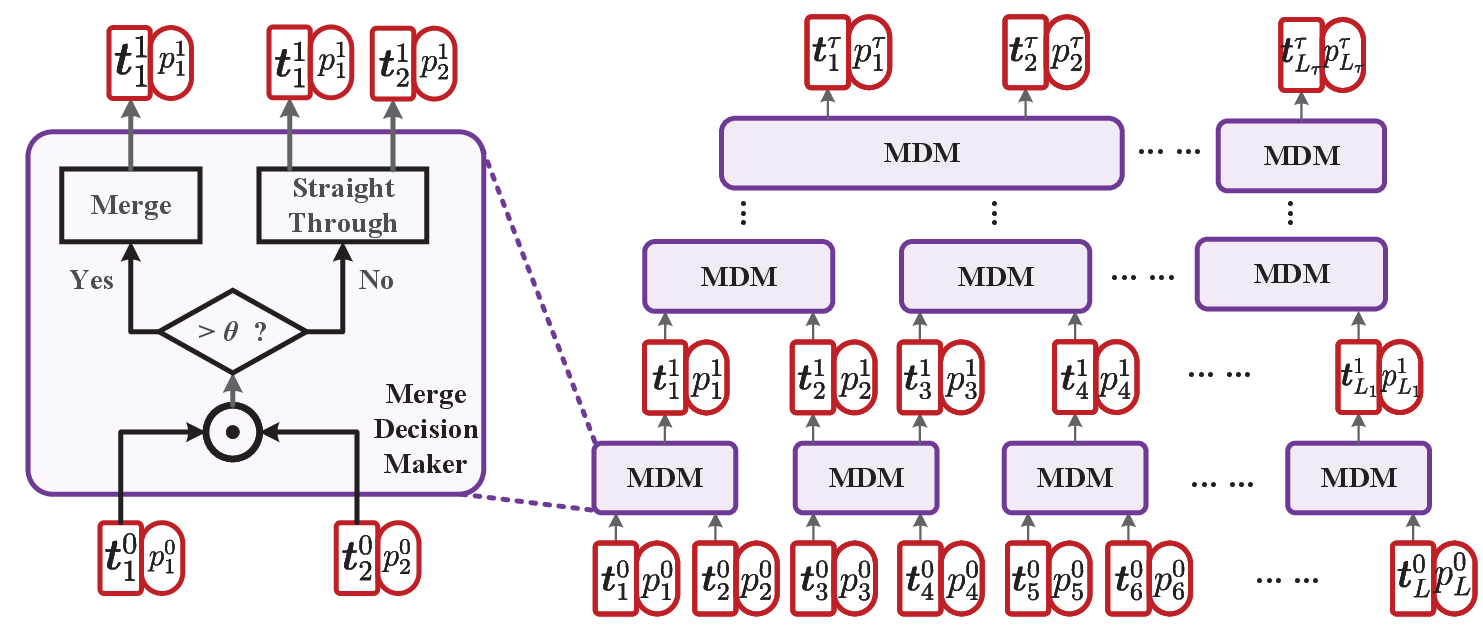}
		\captionsetup{justification=justified, singlelinecheck=false}
		\caption{\small{The adaptive token aggregation (ATA) module. The aggregation is repeated over multiple rounds. In each round, adjacent token pairs are evaluated by a merge decision maker (MDM), which merges tokens with high similarity or passes them through unchanged.}}
	\end{center}
	\label{fig3}
	\vspace{-16pt}
\end{figure*}

\subsection{The Adaptive Token Aggregation (ATA) Module}

The objective of the token aggregation module is to merge the original tokens into a sparse yet informative token sequence, while maintaining adaptive and computationally efficient processing. To fulfill these requirements, we propose a multi-round aggregation algorithm based on channel consistency, as illustrated in Fig. 3. The aggregation procedure is performed over $\tau$ rounds, during which redundant tokens are progressively merged according to their similarity. 

To provide a clear description, the tokens produced after the $i$-th round and their equivalent positions are denoted by $\bm{t}^i_l (1\leqslant l\leqslant L_i)$ and $p^i_l (1\leqslant l\leqslant L_i)$, respectively, where $L_i$ is the number of tokens remaining after the $i$-th round. Following this notation, the initial tokens and their positions are defined as
\begin{equation}
	\bm{t}^0_l=\bm{t}_l,\quad p^0_l=l,\quad(1\leqslant l\leqslant L),
\end{equation}
which correspond to the original token sequence before aggregation.

In each aggregation round, every two neighboring tokens are grouped and processed by a merge decision maker (MDM), which determines whether the two tokens should be merged based on their similarity. As is shown in Fig. 3, we take $\bm{t}^0_1\in\mathbb{C}^{KS\times1}$ and $\bm{t}^0_2\in\mathbb{C}^{KS\times1}$ as the example. In the MDM, the similarity score of the two input tokens is first calculated by a normalized complex inner product as
\begin{equation}
	s^0(1,2)=\mathfrak{R}\!\left\{
	\frac{(\bm{t}^0_1)^H \bm{t}^0_2}
	{\|\bm{t}^0_1\|_2 \, \|\bm{t}^0_2\|_2}
	\right\},
\end{equation}
where $(\cdot)^H$ is the Hermitian transpose, $\|\cdot\|_2$ is the L2-norm of a vector, and $\mathfrak{R}\{\cdot\}$ is the real component of a complex number. This similarity metric captures the alignment degree between two tokens, and thus serves as an effective indicator of channel consistency. The computed similarity score is then compared with a predefined threshold\footnote{The threshold $\theta$ controls the aggressiveness of token merging: a larger $\theta$ results in fewer merges and higher computational cost but better preserved channel information, while a smaller $\theta$ leads to more aggressive compression at the expense of potential performance degradation. In practice, $\theta$ can be selected based on the available computational budget.} $\theta$, resulting in two possible cases as follows. 

\textbf{Case 1:} If $s^0(1,2)>\theta$, the two tokens are considered highly correlated and an element-wise aggregation is applied to produce a merged token $\bm{t}^1_1=\mathcal{M}(\bm{t}^0_1, \bm{t}^0_2)$. Specifically, each element of the merged token, denoted by $\bm{t}^1_1[n] (1\leqslant n\leqslant KS)$, is 
\begin{equation}
	\bm{t}^1_1[n]=\sqrt{|\bm{t}^0_1[n]||\bm{t}^0_2[n]|}\exp\left(j\frac{\arg(\bm{t}^0_1[n])+\arg(\bm{t}^0_2[n])}{2}\right),
	\label{merge1}
\end{equation}
where $\arg(\cdot)$ denotes the phase of a complex number. As shown in (\ref{merge1}), the proposed aggregation operation computes the geometric average between $\bm{t}^0_1$ and $\bm{t}^0_2$. This design preserves the dominant channel characteristics shared by the two input tokens while mitigating noise and small-scale variations. Consequently, the merged token provides a compact yet informative representation that retains the essential channel information contained in $\bm{t}^0_1$ and $\bm{t}^0_2$. The equivalent position is merged as  
\begin{equation}
	p^1_1=\frac{1}{2}(p^0_1+p^0_2).
	\label{merge2}
\end{equation}
Performing equivalent positions merging is crucial because, after aggregation, the resulting token no longer corresponds to a single temporal or frequency sample. Without updating the positional information, the relative physical relationship between tokens would be lost, leading to degraded performance in the subsequent Transformer-based channel feature extraction. With the help of merged equivalent positions, the Transformer blocks are able to correctly model the correlation and structural dependencies among tokens.

\textbf{Case 2:} If $s^0(1,2)\leqslant\theta$, which means that the distance between these two tokens are sufficiently large, merging them would incur non-negligible information loss. In this case, the MDM directly outputs the input tokens without any modifications, i.e.,
\begin{equation}
	\bm{t}^1_1=\bm{t}^0_1,\quad\bm{t}^1_2=\bm{t}^0_2,\quad p^1_1 = p^0_1,\quad p^1_2 = p^0_2.
\end{equation}

The aforementioned decision and aggregation operations are implemented iteratively. After $\tau$ rounds, the output of the ATA module can be obtained as
\begin{equation}
	\tilde{\bm{t}}_l=\bm{t}^\tau_l,\quad p_l = p^\tau_l,\quad (1\leqslant l\leqslant L_\tau).
\end{equation}

It is worth noting that ATA is adaptive since the merge decisions and the resulting sequence length are determined by the input CSI. By merging redundant tokens in smooth channel regions while preserving informative tokens in rapidly varying regions, ATA achieves a favorable trade-off between computational efficiency and channel information preservation.

\subsection{Channel Feature Extractor}

After the ATA module, the aggregated token sequence is fed into the channel feature extractor to produce the feature vectors. The channel feature extractor of ConsisFormer is similar to that of classical Transformer in Fig. 1. The only difference is that a rotary position embedding (RoPE) \cite{ROPE} is utilized in each self-attention head, which is illustrated in Fig. 4. As is shown in Fig. 4, the query and key matrices ($\bm{Q}^{\text{att}}$ and $\bm{K}^{\text{att}}$) are processed before calculating the attention score, by a RoPE module based on equivalent positions to yield
\begin{eqnarray}
	\tilde{\bm{Q}}^{\text{att}}=[\bm{R}(p_1)\bm{q}_1, \bm{R}(p_2)\bm{q}_2, ..., \bm{R}(p_{L_\tau})\bm{q}_{L_\tau}] \in\mathbb{R}^{d\times L_\tau}
	\label{Ropeq}
\end{eqnarray}
and 
\begin{eqnarray}
	\tilde{\bm{K}}^{\text{att}}=[\bm{R}(p_1)\bm{k}_1, \bm{R}(p_2)\bm{k}_2, ..., \bm{R}(p_{L_\tau})\bm{k}_{L_\tau}] \in\mathbb{R}^{d\times L_\tau},
	\label{Ropek}
\end{eqnarray}
where $\bm{q}_l$ and $\bm{k}_l$ are the $l$-th column of $\bm{Q}^{\text{att}}$ and $\bm{K}^{\text{att}}$, respectively. 

\begin{figure}[t]
	\begin{center}
		\includegraphics[scale=0.5]{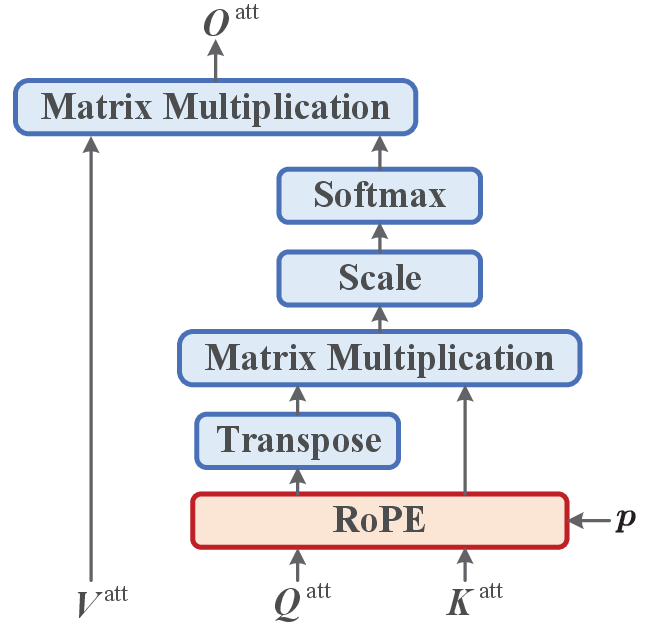}
		\caption{\small{A self-attention head with RoPE.}}
	\end{center}
	\label{fig1}
	\vspace{-22pt}
\end{figure}

In (\ref{Ropeq}) and (\ref{Ropek}), each query or key vector is multiplied by a matrix
\begin{eqnarray}
	\bm{R}(p_{l})=\text{diag}(\bm{R}_1(p_{l}), \bm{R}_2(p_{l}), ..., \bm{R}_{d/2}(p_{l})) \in\mathbb{R}^{d\times d},
	\label{RopeR}
\end{eqnarray}
which denotes a block-diagonal matrix with the $2\times 2$ rotation matrices 
$\bm R_i(p_{l})$ on the diagonal and zero matrices elsewhere. The rotation matrix $\bm R_i(p_{l})$ is constructed based on the equivalent position $p_l$ as 
\begin{eqnarray}
\bm R_i(p_{l}) =
\begin{bmatrix}
	\cos(p_l\theta_i) & -\sin(p_l\theta_i) \\
	\sin(p_l\theta_i) & \cos(p_l\theta_i)
\end{bmatrix},
	\label{RopeRi}
\end{eqnarray}
where $\theta_i=10000^{-\frac{2(i-1)}{d}}$. According to (\ref{RopeR})-(\ref{RopeRi}), the inner product of rotated $l_1$-th query vector and $l_2$-th key vector is 
\begin{eqnarray}
	[\bm{R}(p_{l_1})\bm{q}_{l_1}]^T[\bm{R}(p_{l_2})\bm{k}_{l_2}]=\bm{q}_{l_1}^T \bm{R}(p_{l_2}-p_{l_1}) \bm{k}_{l_2}.
	\label{ip_dif}
\end{eqnarray}

As shown in Eq. (\ref{ip_dif}), RoPE encodes relative positional differences, i.e., $(p_{l_2}-p_{l_1})$, into the attention score, such that the attention between two tokens depends on their relative position rather than absolute positions. This property is particularly beneficial when combined with the proposed ATA module. Although equivalent positions are jointly merged during token aggregation, they represent approximations of the original indices rather than exact values, introducing positional ambiguity that makes absolute positional encoding unreliable \cite{Relative}. Since the internal correlation structure of CSI is primarily governed by relative inter-token positions, RoPE is inherently robust to such approximation errors, ensuring that the aggregated tokens can still accurately capture the essential structural dependencies within each CSI sample.

\subsection{Feature Sequence Interpolation (FSI) Module}

\begin{figure}[!b]
	\vspace{-16pt}
	\begin{center}
		\includegraphics[scale=0.7]{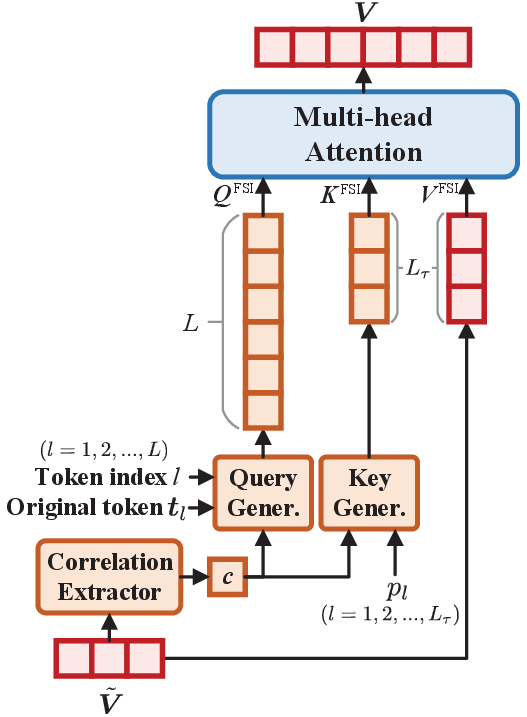}
		\captionsetup{justification=justified, singlelinecheck=false}
		\caption{\small{The structure of the feature sequence interpolation (FSI) module. The module recovers a full-length feature sequence from a sparse one by exploiting the intrinsic correlations among sparse features and the local contextual information provided by the original tokens, through a carefully designed multi-head attention mechanism.}}
	\end{center}
\vspace{-9pt}
\end{figure}

The objective of the FSI module is to recover a full-length feature sequence from the sparse feature representation produced by the channel feature extractor. The overall structure of the FSI module is illustrated in Fig. 5. The core idea of this module is to exploit the intrinsic correlations among the sparse feature vectors, together with the information provided by the original tokens, to perform feature interpolation through a multi-head attention mechanism. To this end, the query, key, and value matrices are carefully designed to serve distinct yet complementary roles in capturing the underlying correlation structure.

Specifically, the sparse feature sequence $\tilde{\bm{V}}=[\tilde{\bm{v}}_1, \tilde{\bm{v}}_2, \dots, \tilde{\bm{v}}_{L_\tau}]$ is first fed into a correlation extractor to infer the global correlation characteristics among the sparse features, which are summarized by a correlation descriptor vector $\bm{c}$. This descriptor provides a compact representation of the internal statistical dependencies within the CSI matrix.

To recover features at all original token positions, query vectors are generated for the entire index set $l=1,2,\dots,L$. At this stage, the original tokens $\bm{t}_l (1\leqslant l\leqslant L)$ are introduced to provide local, position-specific contextual information that may be lost during token sparsification. By incorporating the original tokens together with the correlation descriptor $\bm{c}$ and the token indices $l (1\leqslant l\leqslant L)$, the query generator is able to determine the query vector that needs to be reconstructed for each target position. Specifically, the query vector is generated as
\begin{equation}
	\bm{q}_l^{\text{FSI}} = f_{\theta_1}(\bm{c}, \bm{t}_l, l),
\end{equation}
where $f_{\theta_1}(\cdot)$ denotes the query generation network. With $l = 1,2,\dots,L$, a total of $L$ query vectors are obtained, forming the query matrix $\bm{Q}^{\text{FSI}}$.

Meanwhile, the correlation descriptor $\bm{c}$ and the equivalent position $p_l$ of each sparse feature vector are fed into a key generator to generate the key vector, as
\begin{equation}
	\bm{k}_l^{\text{FSI}} = f_{\theta_2}(\bm{c}, p_l),
\end{equation}
where $f_{\theta_2}(\cdot)$ denotes the key generation network. With $l = 1,2,\dots,L_\tau$, the resulting key vectors form the key matrix $\bm{K}^{\text{FSI}}$. Then, the sparse feature sequence $\tilde{\bm{V}}$ itself is directly used as the value matrix $\bm{V}^{\text{FSI}}$. Finally, $\bm{Q}^{\text{FSI}}$, $\bm{K}^{\text{FSI}}$, and $\bm{V}^{\text{FSI}}$ are processed by a multi-head attention model to finish the interpolation as
\begin{eqnarray}
	\bm{V}=\bm{V}^{\text{FSI}}\sigma(\dfrac{(\bm{Q}^{\text{FSI}})^T\bm{K}^{\text{FSI}}}{\sqrt{d}})\in\mathbb{R}^{d\times L}.
\end{eqnarray}

According to the attention mechanism, the scaled inner product between query and key vectors determines how the sparse feature vectors are weighted and aggregated. By leveraging the correlation descriptor $\bm{c}$ and positional information, the generated queries and keys, the resulting attention weights are explicitly guided by the internal correlation structure of the sparse feature sequence. Consequently, the FSI module can effectively interpolate the sparse feature representation into a full-length sequence, while preserving the intrinsic dependencies within CSI matrix.

Finally, it is worth noting that the original tokens are not involved in the construction of the value matrix. Since the value vectors directly provide the content used for interpolation, incorporating original tokens into value vectors would mix raw input representations with the high-level features extracted by the Transformer blocks, which may degrade the interpolation quality. By restricting the value vectors to the sparse feature sequence, the FSI module ensures that the reconstructed features are derived solely from reliable, correlation-aware representations, while the original tokens affect the reconstruction process only through the query vectors, serving as positional and contextual guidance.

\section{Aggregated Auto-Encoder Pre-training for ConsisFormer}

As a wireless foundation model (WFM) backbone, ConsisFormer is expected to learn task-agnostic channel representations from large-scale CSI data and transfer them to diverse downstream communication and sensing tasks. Therefore, its pre-training is conducted in a self-supervised manner without relying on task-specific labels. In existing WFMs \cite{Boxun}-\!\!\cite{Yucheng1}, masked auto-encoder (MAE) based pre-training \cite{MAE} is widely adopted, as illustrated in Fig. 6(a). In MAE, a subset of input CSI tokens is randomly replaced by constant mask tokens, and the model is trained to reconstruct the masked tokens from the remaining observations. Although this paradigm is effective for generic representation learning, it is not fully aligned with the proposed ConsisFormer architecture, since the ATA module already introduces a structured form of information reduction.

\begin{figure}[t]
	\begin{center}
		\includegraphics[scale=0.5]{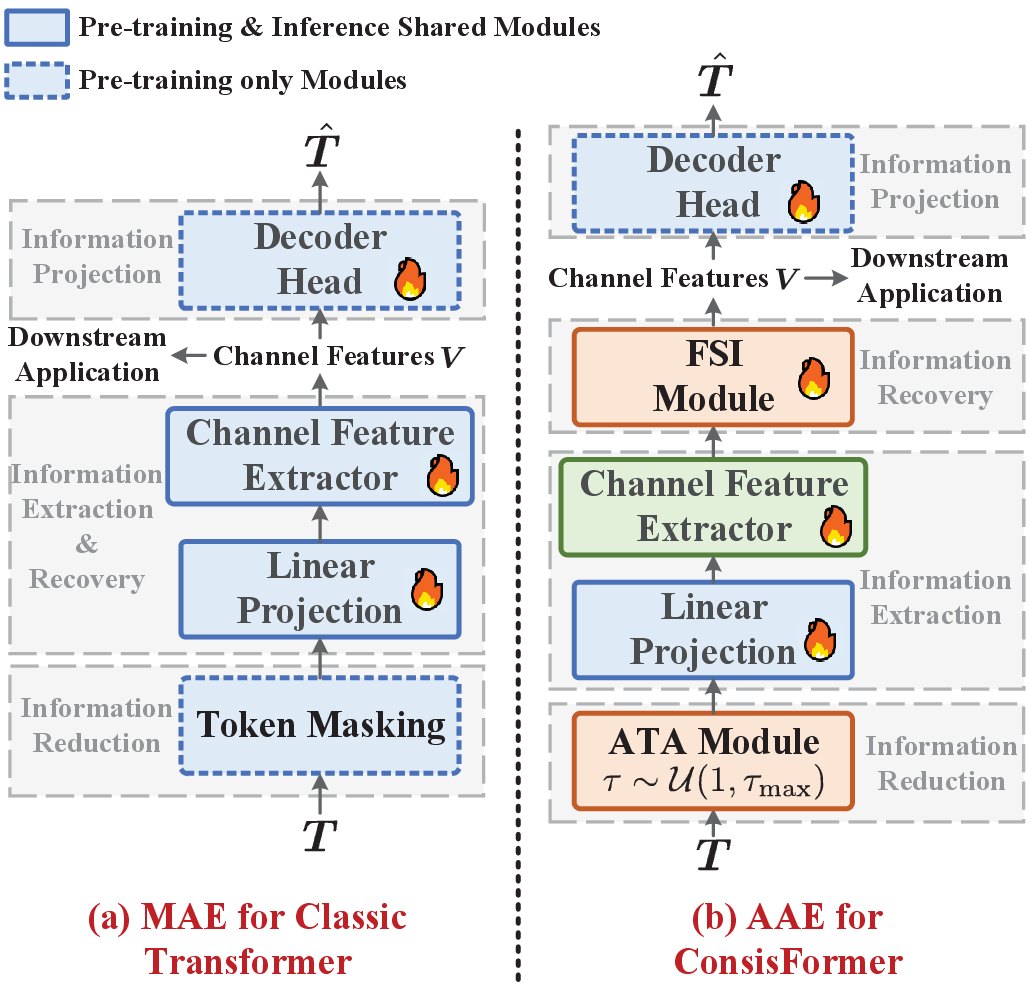}
		\caption{\small{Comparison between MAE and the proposed AAE pre-training paradigm.}}
	\end{center}
	\vspace{-19pt}
\end{figure}

To address this issue, we propose the AAE pre-training paradigm, as shown in Fig. 6(b). The key idea is to replace random token masking with channel consistency-driven token aggregation, such that the pre-training objective is naturally aligned with the inference behavior of the model. Specifically, the ATA module first performs channel-consistency-based aggregation upon the original token sequence $\bm{T}=\{\bm{t}_1,\bm{t}_2,\ldots,\bm{t}_L\}$, producing a compressed token sequence $\tilde{\bm{T}}=\{\tilde{\bm{t}}_1,\tilde{\bm{t}}_2,\ldots,\tilde{\bm{t}}_{L_\tau}\}$, where $L_\tau<L$. During pre-training, the aggregation round $\tau$ is randomly sampled from a discrete uniform distribution over $[1,\tau_{\max}]$, which exposes the model to different token sparsification levels and improves its robustness to varying computational budgets\footnote{At inference time, $\tau$ serves as a flexible hyper-parameter that enables a controllable trade-off between computational complexity and task performance, as will be demonstrated in Section V.}.

The compressed token sequence is then processed by the linear projection and Transformer-based channel feature extractor to obtain sparse channel features. Afterwards, the FSI module recovers a full-length feature sequence $\bm{V}=\{\bm{v}_1,\bm{v}_2,\ldots,\bm{v}_L\}$ based on the sparse feature sequence and the equivalent positional information. Finally, a naive decoder head, usually implemented as a one- or two-layer MLP, maps $\bm{V}$ back to the CSI token domain and reconstructs the original token sequence as $\hat{\bm{T}}$. The network parameters are optimized by minimizing the normalized mean square error (NMSE) between the reconstructed and original tokens, i.e.,
\begin{equation}
	\mathscr{L}_\text{recon}=\text{NMSE}(\hat{\bm{T}},\bm{T}).
\end{equation}

Both MAE and AAE follow the self-supervised principle of recovering complete CSI from partial observations. However, AAE provides two key advantages over MAE. First, MAE randomly masks tokens, risking the loss of critical channel information and introducing physically unrecoverable samples that hinder representation learning. AAE instead exploits channel consistency to aggregate redundant neighboring tokens, ensuring that the retained tokens preserve the essential degrees of freedom of the channel. This leads to a more physically meaningful pre-training objective and more effective channel representations for downstream tasks. Second, MAE suffers from a mismatch between pre-training and inference: the input is randomly masked during pre-training but remains intact during inference. In contrast, AAE is naturally aligned with the inference pipeline, since the same aggregation-extraction-recovery process is applied in both stages. As a result, the model learns representations that are directly optimized for its actual operating conditions.

After pre-training, the ConsisFormer backbone is frozen and transferred to downstream tasks by attaching a lightweight task head to the output of the FSI module. Specifically, a regression head is employed for tasks such as channel prediction and localization, while a classification head is used for LoS/NLoS classification and beam prediction. During downstream adaptation, only the task-specific head is trained with the corresponding task loss, whereas the pre-trained backbone remains fixed. This allows ConsisFormer to efficiently adapt to diverse communication and sensing tasks with minimal training overhead.

\section{Simulation Results and Discussions}

\subsection{Simulation Setups and Baseline Schemes}

\begin{table*}[b] 
	\centering
	\vspace{-5pt}
	\caption{Simulation Settings and Model Configurations}
	\label{tab:simulation_setup}
	\renewcommand{\arraystretch}{1.15}
	\setlength{\tabcolsep}{4pt}
	\begin{tabular}{l|cccc|c}
		\toprule
		& \multicolumn{4}{c|}{\textbf{Wireless Foundation Model (WFM)}}
		& \multirow{2}{*}{\makecell{\textbf{Task-Specific Model}\\\textbf{(Time Domain Ch. Pred.)}}} \\
		\cline{2-5}
		& \textbf{LoS/NLoS Class.} 
		& \textbf{Freq. Domain Ch. Pred.} 
		& \textbf{Beam Pred.} 
		& \textbf{Localization}
		& \\
		
		\midrule
		
		\multicolumn{6}{l}{\cellcolor{gray!15}\textbf{Simulation Settings}} \\
		\midrule
		Sub-carriers Number ($K$)                 
		& 128 & 160 (128 predict 32) & 128 & 128 & 16 \\
		Sub-carrier Spacing                  
		& 15 kHz & 15 kHz & 15 kHz & 15 kHz & 30 kHz \\
		Spatial Ch. Number ($S$)           
		& 32 & 32 & 32 & 32 & 16 \\
		Number of OFDM Symbols ($T$)           
		& 1 & 1 & 1 & 1 & 64 (48 predict 16) \\
		Evaluation Scenarios                         
		& Indianapolis 
		& San Francisco
		& Seattle
		& Outdoor1
		& \makecell{Indianapolis} \\
		\midrule
		
		\multicolumn{6}{l}{\cellcolor{gray!15}\textbf{Model Configuration}} \\
		\midrule
		Tokenization Domain            
		& Frequency & Frequency & Frequency & Frequency & Time \\
		Token Sequence Length ($L$)            
		& 128 & 128 & 128 & 128 & 48 \\
		Aggregation Threshold ($\theta$)         
		& 0.85 & 0.85 & 0.85 & 0.85 & 0.95 \\
		Number of Encoder Blocks ($N_e$)              
		& 16 & 16 & 16 & 16 & 4 \\
		Number of Attention Heads              
		& 8 & 8 & 8 & 8 & 8 \\
		Embedding Dimension          
		& 256 & 256 & 256 & 256 & 512 \\
		Hidden / FFN Dimension          
		& 512 & 512 & 512 & 512 & 1024 \\
		Number of Parameters              
		& 9.59M & 9.59M & 9.59M & 9.59M & 12.41M \\
		Performance Metric                  
		& F1-score & NMSE & F1-score & MEE & NMSE \\
		Task-specific Head       
		& \makecell{MLP\\(5 layers)}
		& \makecell{Transformer Encoder\\(2 layers)} 
		& \makecell{MLP\\(5 layers)}
		& \makecell{MLP\\(4 layers)}
		& \makecell{Transformer Encoder\\(1 layer)} \\
		\bottomrule
	\end{tabular}
\end{table*}

In this section, the open-source wireless channel generator DeepMIMO \cite{Alkhateeb} is employed to generate the CSI datasets. A total of 20 propagation scenarios are considered, among which 16 scenarios are used for pre-training and 4 scenarios are reserved for downstream evaluation. Specifically, the \emph{ASU Campus} scenario and \emph{City0--City14} are used for pre-training, while \emph{City15--City17} and \emph{Outdoor1} are used for evaluation. For each scenario, CSI samples from 2000 users and 3 base stations are generated, resulting in 6000 samples per scenario. The detailed simulation settings and model configurations are summarized in Table I.

The following methods are considered for comparison. For the main results, three methods are compared:

\textbf{1) Transformer}: The classical Transformer architecture with standard positional encoding, pre-trained using MAE.

\textbf{2) RoPE}: Transformer with rotary position embedding, pre-trained using MAE.

\textbf{3) ConsisFormer}: The proposed ConsisFormer architecture pre-trained using the proposed AAE method.

To further investigate the individual contributions of the proposed modules, ablation studies are conducted with the following additional methods:

\textbf{4) RoPE-Random}: Built upon RoPE, tokens are randomly sampled with a sampling ratio of $p_s$ to reduce sequence length before Transformer blocks.

\textbf{5) RoPE-Uniform}: Built upon RoPE, tokens are uniformly sampled at fixed intervals with a sampling ratio of $p_s$.

\textbf{6) RoPE-ATA}: Built upon RoPE, the proposed ATA module is used without full-length feature recovery by FSI.

The comparison among RoPE-Random, RoPE-Uniform, and RoPE-ATA validates the superiority of channel consistency-driven token aggregation over naive token reduction strategies, while the performance gap between RoPE-ATA and ConsisFormer demonstrates the contribution of the FSI module in recovering task-relevant information from the sparsified token sequence.

During downstream evaluation, the task head is trained with $4500\cdot R_t$ samples and tested on a set of $1500$ samples, where $R_t$ denotes the training ratio. Four downstream tasks are considered, briefly introduced as follows.

\textbf{1) LoS/NLoS classification}: In this task, the LoS/NLoS status of the environment is estimated based on the channel representation. The cross entropy is used as loss function for the downstream training, and F1-score is served as the evaluation metric. The F1-score balances precision and recall into a number between 0 and 1, where a higher value indicates better model performance.

\textbf{2) Frequency-domain channel prediction}: This task involves predicting the CSI of 32 adjacent subcarriers based on the channel representations of the existing CSI over 128 subcarriers. The Normalized Mean Squared Error (NMSE) is used both as the loss function and the evaluation metric.

\textbf{3) Beam prediction}: This task utilizes the CSI features of Sub-6GHz (3.5GHz) channels to predict the optimal mmWave (28GHz) beam from a codebook of size 16. Cross-entropy is used as the loss function, while the F1-score serves as the evaluation metric.

\textbf{4) Outdoor localization}: In this task, the location of the UE is estimated based on the features of the CSI between the BS and the UE. In this task, the mean Euclidean error (MEE) between the estimated location and the real location is used as the loss function and evaluation metric.

The computational cost of the NN models is measured in terms of floating-point operations (FLOPs), defined as the number of floating-point operations required to process a single input sample. One giga floating-point operation (GFLOP) equals $10^{9}$ FLOPs. For the baseline methods, the computational cost includes the linear projection and the channel feature extractor. For the proposed ConsisFormer, the costs of the ATA module, the FSI module, the linear projection, and the channel feature extractor are taken into account. Since the task heads are the same across all methods, their computational cost is omitted for comparison purpose.

Further, although ConsisFormer is proposed in this work as the backbone for WFMs, the aggregation–extraction–recovery design is universal. It is not only applicable to WFMs but also to task-specific models. As long as a model is based on the Transformer architecture, the techniques proposed in this paper can be used to achieve efficient computation. To validate this, an additional experiment is conducted where ConsisFormer is directly trained for a specific task: time-domain channel prediction. The task involves predicting the CSI of the subsequent 16 OFDM symbols from historical observations of 48 OFDM symbols, with NMSE as both the training loss and evaluation metric.

\vspace{-0pt}
\subsection{Simulation Results and Discussions for WFMs}

\begin{figure}[t]
		\vspace{-10pt}
	\begin{center}
		\includegraphics[scale=0.55]{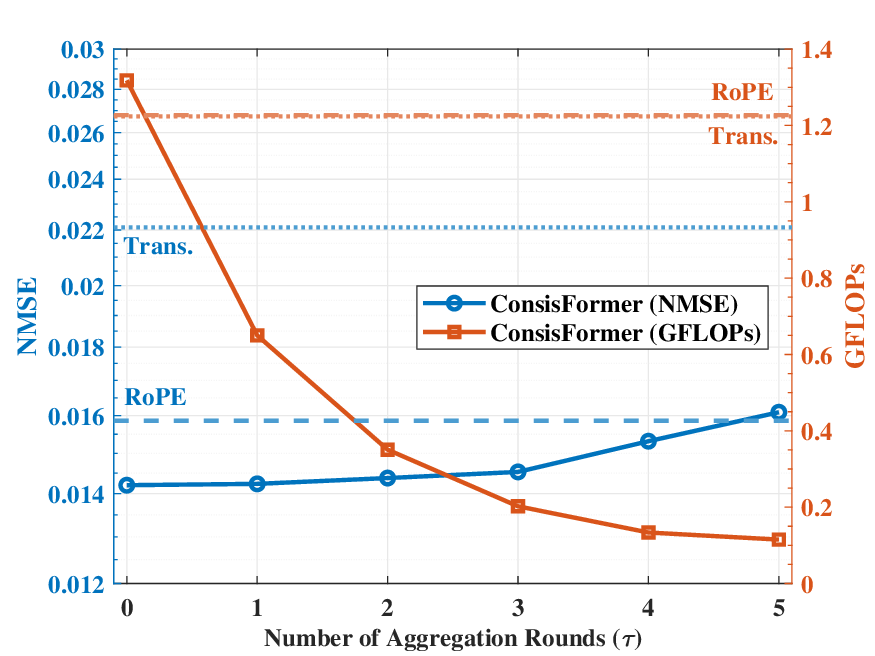}
		\caption{\small{The NMSE and GFLOPs results for channel prediction.}}
	\end{center}
	\vspace{-16pt}
\end{figure}

	Fig. 7 presents the NMSE performance and computational cost of the classical Transformer, RoPE-based Transformer, and ConsisFormer under different aggregation rounds $\tau$ for frequency-domain channel prediction. In Fig. 7, the blue lines represent the performances while the red lines correspond to the costs. When $\tau=0$, ConsisFormer achieves lower NMSE than both the classical Transformer and RoPE baselines. Since the token sequence is not aggregated in this case, this performance gain primarily stems from the proposed AAE pre-training strategy, which encourage the model to learn more informative and physically meaningful channel representations than MAE-based pre-training. As $\tau$ increases, the computational cost of ConsisFormer decreases rapidly, while the NMSE remains almost unchanged for $\tau \leqslant 3$. Specifically, increasing $\tau$ from 0 to 3 reduces the GFLOPs by $84.6\%$, with only a $2.3\%$ NMSE increase. This demonstrates that ATA can effectively remove redundant CSI tokens by exploiting channel consistency, and FSI helps preserve fine-grained channel representations. When $\tau=3$, ConsisFormer reduces the GFLOPs by about $83.5\%$ compared with both baselines, while lowering the NMSE by $34.25\%$ and $8.41\%$ compared with the classical Transformer and RoPE, respectively. Even at $\tau=5$, ConsisFormer still maintains comparable accuracy to RoPE while reducing the computational cost by more than $90\%$, demonstrating a favorable accuracy-complexity trade-off.

\begin{figure}[t!]
	\begin{center}
		\vspace{-8pt}
		\includegraphics[scale=0.55]{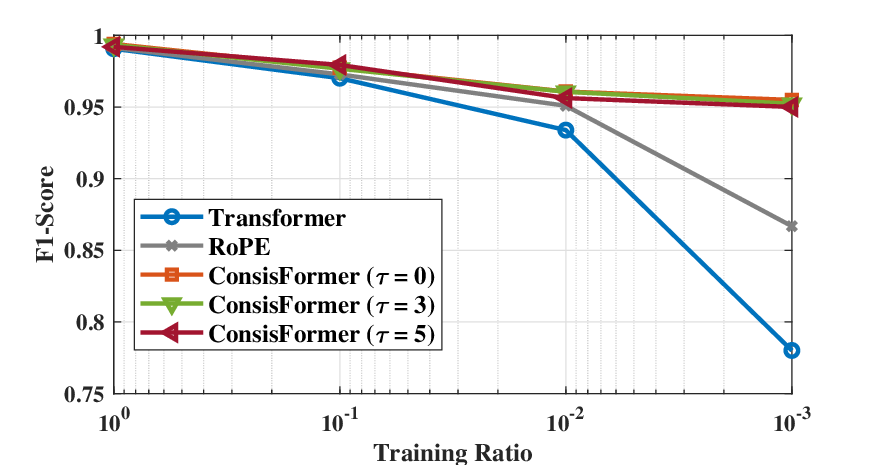}
		\caption{\small{F1-score versus training ratio for Los/NLoS classification.}}
	\end{center}
	\vspace{-22pt}
\end{figure}

 \begin{table}[b]
	\vspace{-10pt}
	\begin{center}
		\caption{{Computational Cost Comparison (GFLOPs)}}
		\begin{tabular}{ccccc}
			\toprule
			Methods & Total & LP $\&$ CFE & ATA& FSI\\
			\midrule
			Transformer & 1.223  & 1.223 & - & - \\
			RoPE & 1.227  & 1.227 & - & - \\
			ConsisFormer ($\tau=0$) & 1.318  & 1.227 & 0 & 0.091 \\
			%	ConsisFormer ($\tau=1$) & 0.650  & 0.580 & 1$\times10^{-5}$ & 0.070 \\
			%	ConsisFormer ($\tau=2$) & 0.351  & 0.291 & 2$\times10^{-5}$ & 0.060 \\
			ConsisFormer ($\tau=3$) & 0.202  & 0.148 & 2$\times10^{-5}$ & 0.055 \\
			%	ConsisFormer ($\tau=4$) & 0.133  & 0.081 & 2$\times10^{-5}$ & 0.052 \\
			ConsisFormer ($\tau=5$) & 0.115  & 0.063 & 2$\times10^{-5}$ & 0.051 \\
			\bottomrule
		\end{tabular}
	\end{center}
\vspace{-6pt}
\end{table} 

Fig. 8 shows the F1-score performance of the LoS/NLoS classification task under different training ratios $R_t$, and the corresponding computational costs are summarized in Table II. As expected, the F1-score decreases with reduced training ratio, since limited training data increases the risk of over-fitting. It can be observed that the proposed ConsisFormer consistently outperforms both baselines, especially for low training ratios. This performance gain demonstrates the advantage of the proposed AAE pre-training strategy, which enables the model to learn more robust channel representations by explicitly exploiting channel consistency, thereby reducing the risk of over-fitting under the situation of data scarcity. Meanwhile, Table II shows that when $\tau=5$, the total computational cost is reduced from $1.318$ GFLOPs ($\tau=0$) to $0.115$ GFLOPs, corresponding to a reduction of $91.27\%$, while the F1-score remains almost unchanged compared with the case of $\tau=0$. This observation is consistent with the fact that LoS/NLoS classification relies mainly on coarse channel characteristics and is therefore insensitive to aggressive token aggregation. Even with a large number of aggregation rounds, the essential discriminative features are well preserved, enabling significant complexity reduction with negligible performance degradation.

\begin{figure}[t]
	\vspace{-0pt}
	\begin{center}
		\includegraphics[scale=0.6]{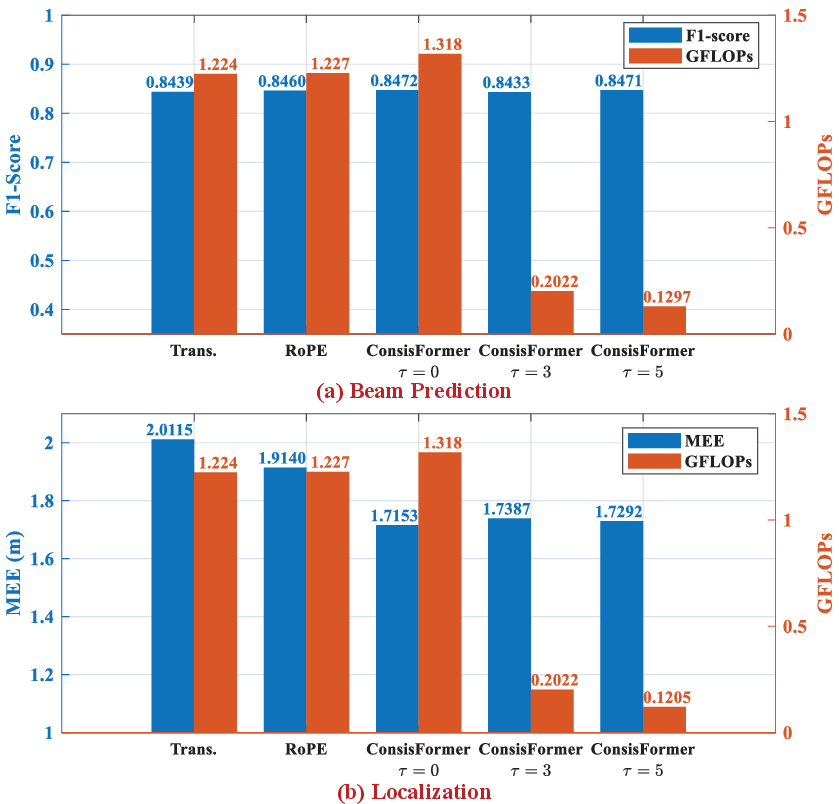}
		\caption{\small{Simulation results for beam prediction and localization.}}
	\end{center}
	\vspace{-20pt}
\end{figure}

Fig. 9 exhibits the performance of two downstream tasks, i.e., beam prediction and localization, together with their corresponding computational costs. For both tasks, the proposed ConsisFormer achieves competitive or improved performance compared with the baseline models, while significantly reducing the computational cost when token aggregation is enabled. For the beam prediction task shown in Fig. 9(a), all models exhibit very close F1-scores, regardless of the model architecture or the aggregation round $\tau$. This phenomenon can be attributed to the inherent challenges of cross-band beam prediction, specifically from 3.5GHz channels to 28GHz mmWave beams. Due to the limited overlap in scattering characteristics between these widely separated frequency bands, the achievable prediction accuracy is fundamentally constrained. In contrast, the localization task in Fig. 9(b) exhibits more pronounced performance variations across different models. Compared with the classical Transformer and RoPE, ConsisFormer achieves lower localization error, indicating that the learned channel representations preserve more fine-grained spatial information. Even with aggressive aggregation ($\tau=5$), the localization performance degradation remains limited, suggesting that the proposed ATA and FSI mechanisms effectively retain the essential spatial information required for localization. 

\vspace{-12pt}
\subsection{Ablation Study}

\begin{figure}[t]
	\vspace{-0pt}
	\begin{center}
		\includegraphics[scale=0.6]{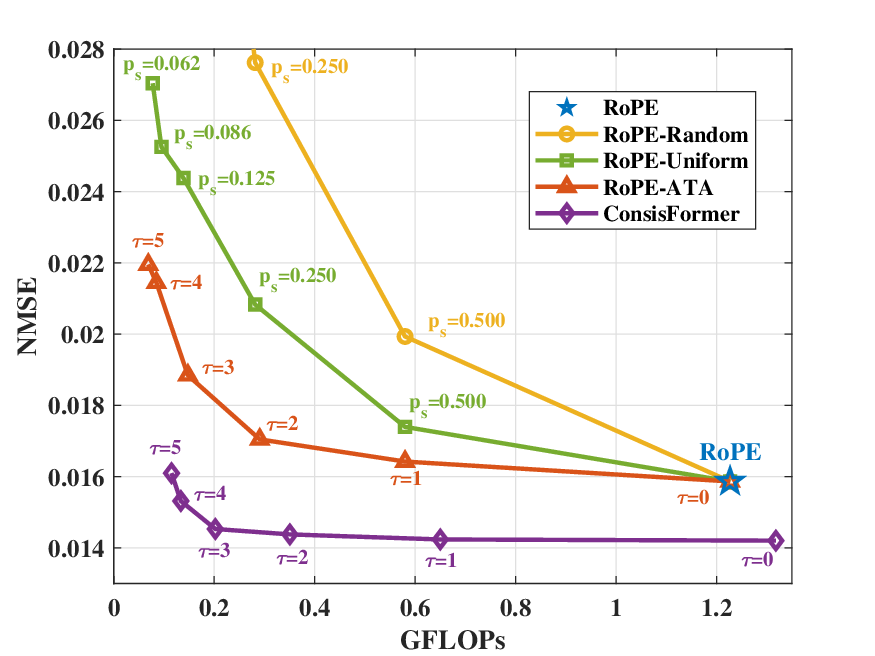}
		\caption{\small{The ablation study on channel prediction.}}
	\end{center}
	\vspace{-22pt}
\end{figure}

\begin{figure}[b]
	\vspace{-18pt}
	\begin{center}
		\includegraphics[scale=0.55]{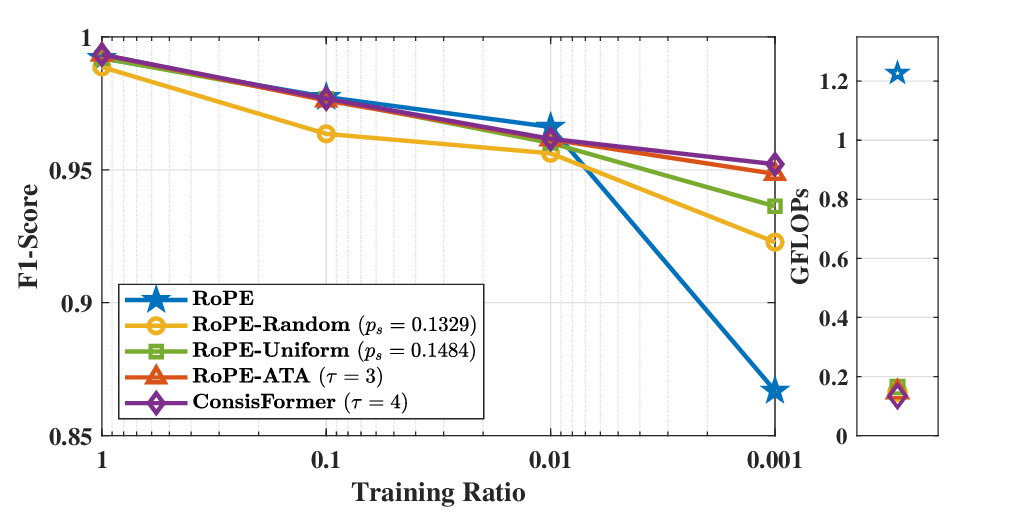}
		\caption{\small{The ablation study on Los/NLoS classification.}}
	\end{center}
	\vspace{-6pt}
\end{figure}

Fig. 10 presents the NMSE versus GFLOPs trade-off curves for the ablation study on channel prediction, where the points closer to the lower-left corner are more favorable, indicating both lower computational cost and smaller prediction error. For the ATA-based methods, different points are obtained by varying the aggregation iteration number $\tau$, while for random and uniform sampling, the curves are obtained by adjusting the sampling ratio $p_s$. The random and uniform sampling methods reduce the computational cost by discarding tokens, but they suffer from significant performance degradation as the sampling ratio decreases. In contrast, the proposed ATA method effectively alleviates this degradation by exploiting channel consistency for structured token aggregation. Furthermore, ConsisFormer consistently outperforms RoPE-ATA across all computational costs by a significant margin, validating the contribution of the FSI module in recovering task-relevant information from the sparsified token sequence.

Fig. 11 shows the ablation results for LoS/NLoS classification under different training ratios, with the left panel showing the F1-score and the right panel reporting the corresponding GFLOPs. Unlike channel prediction, this task highlights the generalization behavior under limited labeled data. Interestingly, when $R_t=0.001$, token sparsification methods outperform the RoPE baseline, suggesting that removing redundant CSI details can reduce over-fitting of the downstream classifier. Among the sparsification schemes, RoPE-ATA performs best by retaining structurally informative channel variations. ConsisFormer achieves the highest F1-score overall, indicating that FSI further improves generalization ability of the task head.

\vspace{-6pt}
\subsection{Visualization Results}

\begin{figure}[t]
	\vspace{-0pt}
	\begin{center}
		\includegraphics[scale=0.65]{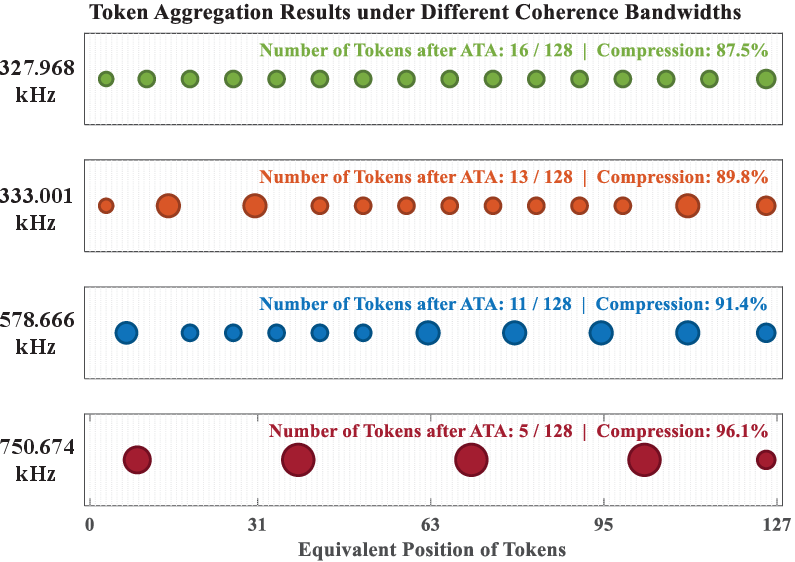}
		\caption{\small{The token aggregation results of different samples.}}
	\end{center}
	\vspace{-22pt}
\end{figure}

\begin{figure}[b]
	\vspace{-12pt}
	\begin{center}
		\includegraphics[scale=0.29]{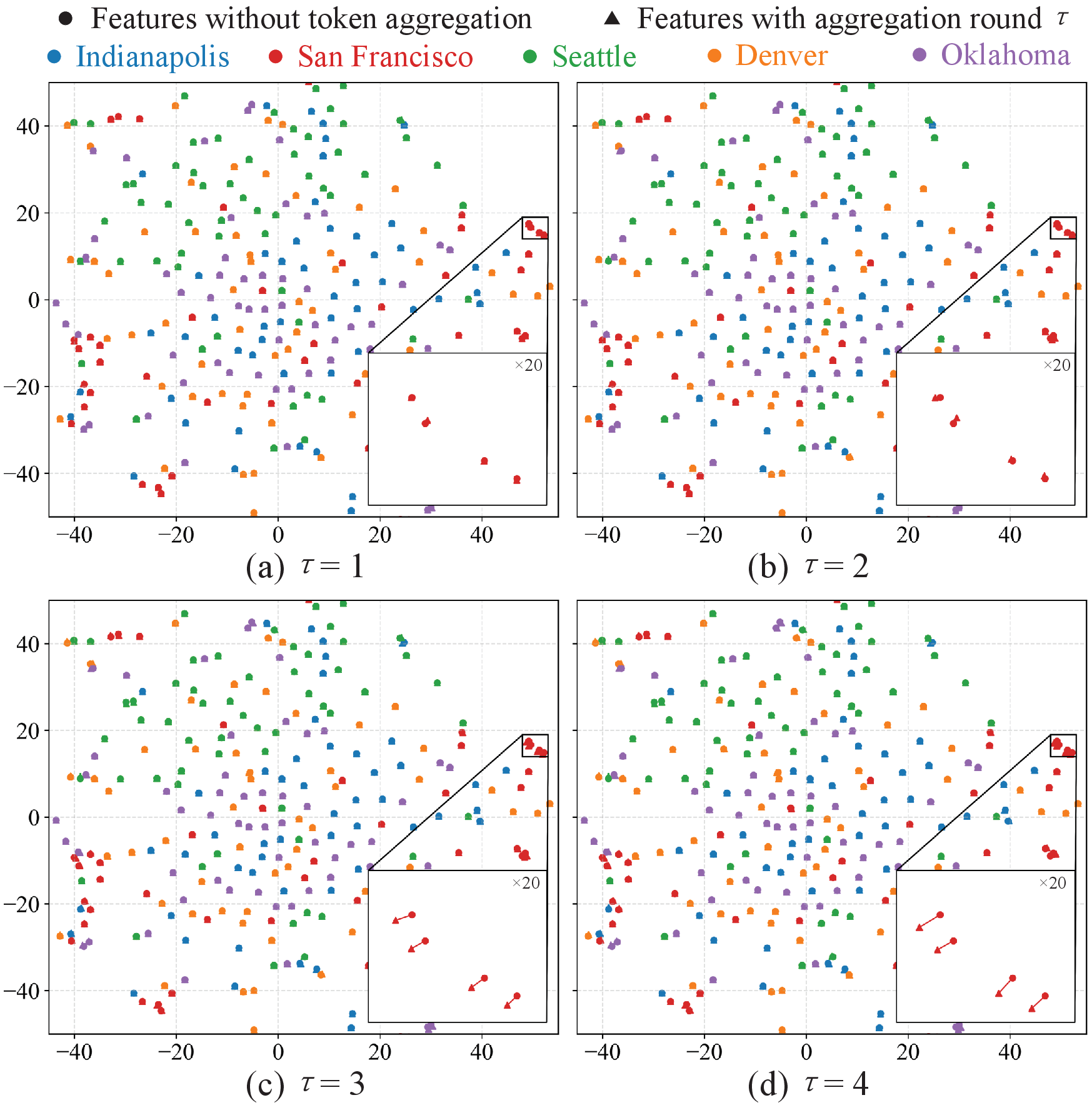}
		\caption{\small{Channel feature visualization based on t-SNE.}}
	\end{center}
	\vspace{-6pt}
\end{figure}

Fig. 12 visualizes the token aggregation results of ATA when $\tau=5$ for channel samples with different coherence bandwidths. The horizontal axis denotes the equivalent position of each retained token after aggregation, while the marker size represents the number of original tokens represented by that aggregated token. It can be observed that the aggregation patterns vary with the channel coherence bandwidth. As the coherence bandwidth increases, stronger frequency-domain channel consistency allows more neighboring tokens to be merged, leading to fewer retained tokens and higher compression ratios. For example, the number of tokens after ATA decreases from 16 to 5 as the coherence bandwidth increases from 327.968 kHz to 750.674 kHz. Moreover, the retained tokens are not uniformly distributed, and different aggregated tokens represent different numbers of original tokens even within the same channel sample. This indicates that ATA does not perform a fixed or uniform token reduction, but adapts its aggregation behavior to the local correlation structure of the channel. These results provide an intuitive illustration that the proposed ATA is driven by the propagation characteristics of wireless channels, rather than being a generic token sparsification trick.

We further examine whether ConsisFormer preserves informative channel representations after token aggregation. To this end, the high-dimensional channel features produced by ConsisFormer with different aggregation depths $\tau$ are projected onto a two-dimensional space using t-distributed stochastic neighbor embedding (t-SNE) \cite{Maaten}. In Fig. 13, the circles represent the channel features generated by ConsisFormer without token aggregation, while the triangles denote the features obtained with token aggregation under different values of $\tau$. We observe that for all the CSI samples, the projected features before and after token aggregation almost completely overlap. To further examine subtle differences, we select several samples with the largest separation and present a $20\times$ zoom-in view in the inset. Even under this magnified view, only minor displacements can be observed for $\tau=3$ and $\tau=4$, while the feature distributions remain highly consistent. These results suggest that the proposed ATA module effectively removes redundant information without altering the essential channel characteristics, and the FSI module successfully compensates for the information loss induced by aggregation.

\vspace{-6pt}
\subsection{Simulation Results for Task-specific Models}

\begin{figure}[b]
	\vspace{-22pt}
	\begin{center}
		\includegraphics[scale=0.57]{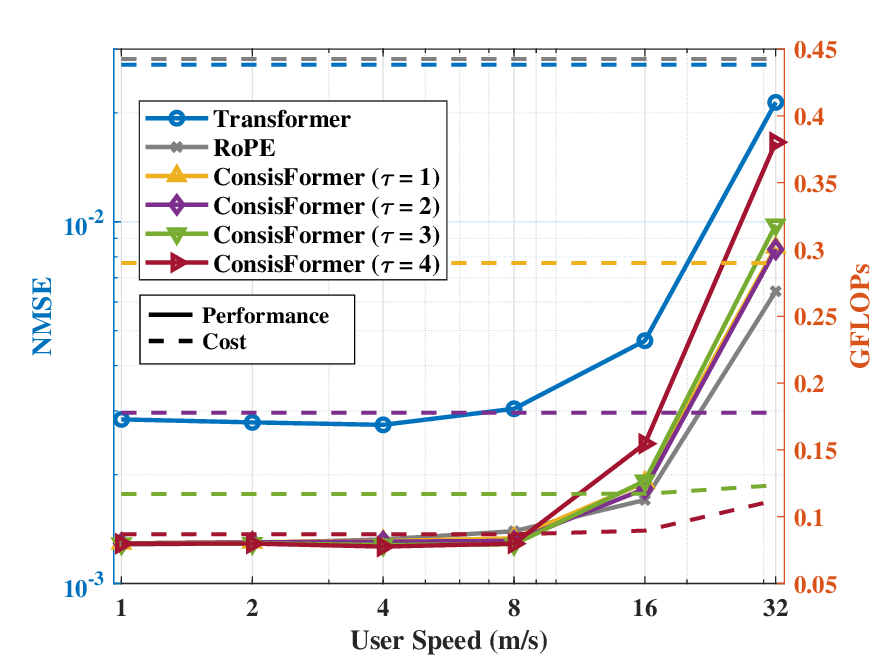}
		\caption{\small{The simulation results under various user speeds.}}
	\end{center}
	\vspace{-6pt}
\end{figure}

Fig. 14 illustrates the results of different Transformer-based structures when they are deployed as task-specific (time-domain channel prediction) models, evaluated under various user speeds ranging from 1 m/s to 32 m/s. The solid curves denote the NMSE and the dashed curves represent the corresponding GFLOPs. Compared with the classical Transformer and RoPE, the proposed ConsisFormer achieves significantly lower computational cost while maintaining competitive prediction accuracy across different aggregation rounds $\tau$, demonstrating the effectiveness of the proposed ConsisFormer as a task-specific model.

As the user speed increases, the NMSE of all methods increases, for the reason that faster channel variations make channel prediction more challenging. It can also be observed that the performance gap between ConsisFormer and RoPE becomes larger in the high-mobility regime, especially when a larger aggregation round $\tau$ is used. This is reasonable and as expected, since high mobility weakens temporal channel consistency and makes fine-grained temporal variations more important. In this case, aggressive token aggregation may inevitably remove part of the rapidly varying channel information. Nevertheless, ConsisFormer still provides a favorable trade-off between prediction accuracy and computational cost.

\section{Conclusions and Future Works}

In this paper, we have proposed ConsisFormer, a compute-efficient Transformer architecture intrinsically motivated by channel consistency. By explicitly reducing redundant input tokens and recovering complete channel information through structured feature correlations, ConsisFormer is able to generate compact yet informative channel features with significantly reduced computational cost. Extensive simulation results have demonstrated that, compared with the existing structures, ConsisFormer consistently achieves superior performance with substantially lower complexity for diverse downstream tasks. For tasks that only require coarse channel features such as LoS/NLoS classification and beam prediction, ConsisFormer reduces the computational cost by more than $91\%$ without noticeable performance degradation. Even for tasks that need fine-grained channel representations, including channel prediction, the proposed model still achieves over $83\%$ reduction in GFLOPs while maintaining comparable prediction accuracy.

Future work will explore the adaptive selection of aggregation hyper-parameters, including the aggregation round and threshold, by learning the mapping from channel conditions and computational budgets to their appropriate values. Another promising direction is to validate ConsisFormer in practical wireless systems and extend it to more dynamic propagation environments.

	\ifCLASSOPTIONcaptionsoff
	\newpage
	\fi


\vspace{-10pt}
	\begin{thebibliography}{99}
\bibitem{ITU} ITU-R M.2160-0, \emph{Framework and overall objectives of the future development of IMT for 2030 and beyond}, ITU, Nov. 2023.

\bibitem{Jun} C. Wen, J. Tong, Y. Hu, Z. Lin, and J. Zhang, ``Neural representation for wireless radiation field reconstruction: A 3D Gaussian splatting approach,'' \textit{IEEE Trans. Wireless Commun.}, early access, Nov. 2025.

\bibitem{Liangle} H. Zhang, K. Huang, C. Yang, L. Liang, C. Guo, and H. Ye, ``Fast adversarial training for graph neural network based resource allocation,'' \textit{IEEE Wireless Commun. Lett.}, vol. 15, pp. 126-130, Oct. 2025.

\bibitem{Zhaoyang} Y. Yang, O. Alhussein, A. Arani, Z. Zhang, and M. Debbah, ``Generative diffusion receivers: Achieving pilot-efficient MIMO-OFDM,'' \emph{arXiv preprint arXiv:2506.18419}, Jun. 2025.

\bibitem{JSAC} L. Sun, Y. Wang, A. L. Swindlehurst, and X. Tang, ``Generative-adversarial-network enabled signal detection for communication systems with unknown channel models,'' \textit{IEEE J. Select. Areas Commun.}, vol. 39, no. 1, pp. 47-60, Jan. 2021.

\bibitem{Wang} P. Wang, Z. Wang, and Z. Han, ``Intelligent wireless interference identification with lightweight Transformer network,'' \textit{IEEE Trans. Commun.}, vol. 73, no. 11, pp. 11355-11367, Nov. 2025.

\bibitem{Zhang} S. Zhang, S. Zhang, W. Yuan, and T. Q. S. Quek, ``Transformer-empowered predictive beamforming for rate-splitting multiple access in non-terrestrial networks,'' \textit{IEEE Trans. Wireless Commun.}, vol. 23, no. 12, pp. 19776-19788, Dec. 2024.

\bibitem{Chafaa} I. Chafaa, G. Bacci, and L. Sanguinetti, ``Transformer-based power optimization for max-min fairness in cell-free massive MIMO,'' \textit{IEEE Wireless Commun. Lett.}, vol. 14, no. 8, pp. 2316-2320, Aug. 2025.

\bibitem{Zhou} T. Zhou, X. Liu, Z. Xiang, H. Zhang, B. Ai, and L. Liu, ``Transformer network based channel prediction for CSI feedback enhancement in AI-native interface,'' \textit{IEEE Trans. Wireless Commun.}, vol. 23, no. 9, pp. 11154-11167, Sep. 2024.

\bibitem{Chen} L. Chen and L. Sun, ``Self-attention-based real-time signal detector for communication systems with unknown channel models,'' \textit{IEEE Commun. Lett.}, vol. 25, no. 8, pp. 2639-2643, Aug. 2021.

\bibitem{Liu} B. Liu, X. Liu, S. Gao, X. Cheng, and L. Yang, ``LLM4CP: Adapting large language models for channel prediction,'' \textit{J. Commun. Inf. Netw.}, vol. 19, no. 2, pp.113-125, Jun. 2024.

\bibitem{Yucheng} Y. Sheng, K. Huang, L. Liang, P. Liu, S. Jin, and Y. Li, ``Beam prediction based on large language models,'' \emph{arXiv preprint arXiv:2408.08707}, Feb. 2025.

\bibitem{Yizhu} Y. Zhu, L. Yu, L. Shi, J. Zhang, and G. Liu, ``Multi-modal large models based beam prediction: An example empowered by DeepSeek,'' \emph{arXiv preprint arXiv:2506.0592}, Jun. 2025.

\bibitem{Boxun} B. Liu, S. Gao, X. Liu, X. Cheng, and L. Yang, ``WiFo: Wireless foundation model for channel prediction,'' \textit{Sci. China Inf. Sci.}, vol. 68, no. 6, Jun. 2025.

\bibitem{Feifei2} B. Lin, H. Zhang, Y. Jiang, Y. Wang, T. Zhang, S. Yan, H. Li, Y. Liu, and F. Gao, ``A generative pre-trained language model for channel prediction in wireless communications systems,'' in \textit{Proc. 2025 Conf. Empirical Methods Natural Language Process.}, Suzhou, China, Nov. 2025, pp. 13417-13430.

\bibitem{Sadjad} S. Alikhani, G. Charan, and A. Alkhateeb, ``Large wireless model (LWM): A foundation model for wireless channels,'' \emph{arXiv preprint arXiv:2411.08872}, Apr. 2025.

\bibitem{Yang} T. Yang, P. Zhang, M. Zheng, Y. Shi, L. Jing, J. Huang, and N. Li, ``WirelessGPT: A generative pre-trained multi-task learning framework for wireless communication,'' \textit{IEEE Net.}, vol. 39, no. 5, pp. 58-65, Sep. 2025.

\bibitem{Liwen} L. Jing, T. Yang, H. Zhang, Y. Shi, C. Zhang, and B. Zhang, ``Signal compression for wireless communication and sensing: A general approach utilizing pretrained wireless foundation models'' \textit{IEEE Trans. Mob. Comput.}, early access, Mar. 2026.

\bibitem{Yucheng1} Y. Sheng, J. Wang, X. Zhou, L. Liang, H. Ye, S. Jin, and Y. Li, ``A wireless foundation model for multi-task prediction,'' \emph{arXiv preprint arXiv:2507.05938}, Jul. 2025.

\bibitem{Shugong} J. Jiang, W. Yu, Y. Li, Y. Gao and S. Xu, ``A MIMO wireless channel foundation model via CIR-CSI consistency,'' in \textit{Proc. IEEE Int. Conf. Mach. Learn. Commun. Netw. (ICMLCN)}, Barcelona, Spain, May, 2025, pp. 1-6.

\bibitem{MUSE} T. Zheng, J. Guo, L. Dai, S. Jin, and J. Zhang, ``MUSE-FM: Multi-task environment-aware foundation model for wireless communications,'' \emph{arXiv preprint arXiv:2509.01967}, Sep. 2025.

\bibitem{Vaswani}
A. Vaswani \emph{et al.}, ``Attention is all you need,'' in \textit{Proc. Int. Conf. Neural Inf. Process. Syst. (NIPS)}, Long Beach, CA, USA, Dec. 2017, vol. 30, pp. 5998–6008.

\bibitem{DHou} D. Hou, L. Li, W. Lin, J. Liang, and Z. Han, ``ClST: A convolutional transformer framework for automatic modulation recognition by knowledge distillation,'' \textit{IEEE Trans. Wireless Commun.}, vol. 23, no. 7, pp. 8013–8028, Jul. 2024.

\bibitem{Tinywifo} H. Zhang, S. Gao, and X. Cheng, ``Tiny-WiFo: A lightweight wireless foundation model for channel prediction via multi-component adaptive knowledge distillation," \textit{IEEE Wireless Commun. Lett.}, vol. 15, pp. 1846-1850, Feb. 2026.

\bibitem{Byeon} Y. Byeon, D. Kim, Y. Cao, and W. Lim, ``Pruned and quantized hybrid models for edge-based automatic modulation recognition," \textit{IEEE Internet Things J.}, early access, Apr. 2026.

\bibitem{Bolya} Y. Bolya, C. Fu, X. Dai, P. Zhang, C. Feichtenhofer, and J. Hoffman, ``Token merging: Your ViT but faster,'' in \textit{Proc. Int. Conf. Learn. Represent. (ICLR)}, Kigali, Rwanda, 2023.

\bibitem{Renggli} C. Renggli, A. Pinto, N. Houlsby, B. Mustafa, J. Puigcerver, and C. Riquelme, ``Learning to merge tokens in vision transformers,'' \emph{arXiv preprint arXiv:2202.12015}, Feb. 2022.

\bibitem{FZeng} F. Zeng and D. Yu, ``M2M-TAG: Training-free many-to-many token aggregation for vision transformer acceleration,'' in \textit{Proc. Adv. Neural Inf. Process. Syst. (NeurIPS)}, Vancouver, BC, Canada, 2024.

\bibitem{JYun}
J. Yun, M. Kim, and Y. Kim, ``Focus on the core: Efficient attention via pruned token compression for document classification,'' in \textit{Findings of the Assoc. Comput. Linguist. (EMNLP)}, Singapore, 2023, pp. 13617-13628.

\bibitem{PDong}
P. Dong, H. Zhang, G. Y. Li, I. S. Gaspar, and N. NaderiAlizadeh, ``Deep CNN-based channel estimation for mmWave massive MIMO systems,'' \textit{IEEE J. of Sel. Topics in Signal Process.}, vol. 13, no. 5, pp. 989-1000, Sep. 2019.

\bibitem{PBello}
P. Bello, ``Characterization of randomly time-variant linear channels,'' \textit{IEEE Trans. Commun.}, vol. 11, no. 4, pp. 360-393, Dec. 1963.

\bibitem{ROPE} J. Su, Y. Lu, S. Pan, A. Murtadha, B. Wen, and Y. Liu, ``RoFormer: Enhanced transformer with rotary position embedding," \emph{arXiv preprint arXiv:2104.09864}, Apr. 2021.

\bibitem{Relative} P. Shaw, J. Uszkoreit, A. Vaswani, ``Self-Attention with relative position representations," \emph{arXiv preprint arXiv:1803.02155}, Mar. 2018.

\bibitem{MAE}
K. He, X. Chen, S. Xie, Y. Li, P. Doll\'{a}r, and R. Girshick, ``Masked autoencoders are scalable vision learners,'' in \textit{Proc. IEEE Conf. Comput. Vis. Pattern Recognit. (CVPR)}, New Orleans, LA, USA, 2022.

\bibitem{Alkhateeb} A. Alkhateeb, ``DeeepMIMO: A generic deep learning dataset for millimeter wave and massive MIMO applications,'' in \textit{Proc. IEEE Inf. Theory Appl. Workshop (ITA)}, San Diego, CA, USA, Feb. 2019, pp. 1-8.

\bibitem{Maaten} L. Maaten and G. Hinton, ``Visualizing high-dimensional data using t-SNE," \textit{J. Mach. Learn. Res.}, vol. 9, no. 86, pp. 2579–2605, Nov. 2008.
		
	\end{thebibliography}
\end{document}